\documentclass[authoryear,12pt,3p]{jowarticle}

\usepackage[english]{babel}
\usepackage[utf8]{inputenc}
\usepackage{amsmath}
\usepackage{graphicx}
\usepackage[colorinlistoftodos]{todonotes}
\usepackage{natbib}

\makeatletter
\renewcommand\section{\@startsection{section}{1}{\z@}{-3.25ex plus -1ex minus -.2ex}{1.5ex plus .2ex}{\normalsize\bf}}
\renewcommand\subsection{\@startsection{subsection}{2}{\z@}{-3.25ex plus -1ex minus -.2ex}{1.5ex plus .2ex}{\normalsize\bf}}
\renewcommand\subsubsection{\@startsection{subsubsection}{3}{\z@}{-3.25ex plus -1ex minus -.2ex}{1.5ex plus .2ex}{\normalsize\bf}}
\makeatother

\begin{document}
\begin{frontmatter}
\title{Endogenous Epistemic Factionalization}

\author{James Owen Weatherall}\ead{weatherj@uci.edu} \author{Cailin O'Connor}\ead{cailino@uci.edu}
\address{Department of Logic and Philosophy of Science \\ University of California, Irvine}

\date{\today}

\begin{abstract}
Why do people who disagree about one subject tend to disagree about other subjects as well?  In this paper, we introduce a model to explore this phenomenon of ``epistemic factionization''.  Agents attempt to discover the truth about multiple propositions by testing the world and sharing evidence gathered.  But agents tend to mistrust evidence shared by those who do not hold similar beliefs.  This mistrust leads to the endogenous emergence of factions of agents with multiple, highly correlated, polarized beliefs.
  \end{abstract}
\end{frontmatter}

\section{Introduction}

Why, in the United States, do the same people who believe that climate science is bunk also tend to believe that believe widespread gun ownership poses relatively few risks?\footnote{For evidence that these beliefs are, in fact, correlated, see \citet{Kahan}.}  In other words, why do people form \emph{epistemic factions}?  One common explanation appeals to political ideology. For instance, a libertarian ideology might account for the correlation between these beliefs, insofar as it might lead to skepticism about scientific results that might support policies that impinge on individual liberty.  But then why do the same people also tend to believe that legalizing marijuana would pose an unacceptable social risk?  And what about the fact that the same people also tend to believe that economically disadvantaged Black Americans just need to work harder to achieve economic equality with whites?\footnote{See \citet{benegal2018spillover} for recent empirical work on this phenomenon.  Our inference that such factions exist more generally is based on voting behavior in the United States and correlations between beliefs about matters of scientific consensus and political party identification, as, for instance, in \citet{Gallup}.}  In this case, the beliefs do not seem to be connected by any (coherent) background ideology, but they correlate nonetheless.\footnote{\label{fn:trivial} Of course, one can always define an ideology by a disjunctive procedure, basically by stipulating that the ideology consists in believing precisely those things that members of an epistemic faction believe.  But to do so would be to eliminate all explanatory power of positing an underlying ideology, and so we set this possibility aside.  One could also argue that factions are constructed by explicit political alliance building, and so one should not expect there to be any underlying ideological explanation, aside from broad compatibility of policy goals.  We are sympathetic with this suggestion, but set it aside, because we do not think it completely accounts for the epistemic character of the phenomenon we are discussing.}

This paper analyzes a simple model that provides a different explanation of how epistemic factions form, which does not appeal to underlying ideology.   Our model extends one introduced by \citet{o2018scientific} to study the phenomenon of polarization about matters of fact.\footnote{As we describe below, these models are based on the ``network epistemology'' framework developed by \citet{venkatesh1998learning} and introduced to philosophy of science by \citet{zollman2007communication}.}  By `polarization' we mean a situation in which subgroups in a society hold stable, opposing beliefs, despite the fact that they engage in discussion of the relevant issues.\footnote{In contrast, `belief polarization' often refers to the more limited phenomenon where individuals update their credences in different directions in light of the same evidence \citep{dixit2007political, jern2014belief,benoit2014theory}.  Psychologists sometimes refer to `group polarization' or `attitude polarization', which is the phenomenon where a group of individuals will develop more extreme beliefs as a result of discussion of a topic.  Both of these phenomena likely relate to the larger phenomenon we address here, but they are not the focus of the current study.  \citet{bramson2017understanding} give a nice discussion of the various ways that groups can polarize in our sense.}  As \citet{o2018scientific} show, when individuals discount evidence provided by peers who do not share their beliefs, polarization emerges, despite the fact that individuals seek to learn the truth, and are able to gather and share evidence from the world.  In the present paper we consider what happens in this model when agents simultaneously form beliefs in multiple arenas and make decisions about what evidence to trust based on shared belief across these arenas. We find that  epistemic factions form endogenously, with agents holding correlated, polarized beliefs about unrelated topics.

Although real-world polarization across multiple beliefs seems to depend on many factors that are not represented in the models here, they can nevertheless shed light on the phenomenon of interest.  First, the models make clear how epistemic factionalization can occur even among highly rational actors.  The agents in the model adopt a certain, arguably reasonable, heuristic for evaluating the reliability of evidence; they do not succumb to cognitive biases or motivated reasoning, nor to social pressure.  Second, perhaps more important, is that the models show that unrelated beliefs can become strongly correlated even absent other factors, such as shared ideology, shared economic interest, personal identity, or any similarity between the beliefs at issue.  Indeed, not including these factors is, in our view, a strength of the modeling framework.  By stripping away factors such as ideology and political identity, we can focus attention on which aspects of polarization and epistemic factionalization can emerge solely from trust grounded in shared belief.\footnote{We emphasize that we do not mean to argue that epistemic factions never arise due to a common cause.  Rather, this is an example of ``how possibly'' modeling intended to explore what some minimal conditions for a phenomenon might be.}

The paper will proceed as follows.  Sections \ref{sec:literature} and \ref{sec:modeling} present previous literature related to the phenomenon under consideration. In section \ref{sec:model} we introduce the modeling paradigm employed here and also present the model analyzed in the paper.  We present results in section \ref{sec:results}.  As we show, when trust is grounded in shared belief over multiple arenas, beliefs tend to strongly correlate.  This can happen when new beliefs form among agents who already hold polarized opinions.  It can also happen when agents develop multiple beliefs at the same time.  And it can happen over more than two beliefs.  In addition, as we show, the social mistrust that leads to correlated polarization in these models also generally leads agents to develop less accurate beliefs.   Section \ref{sec:conclusion} concludes.

\section{Polarization and Epistemic Factions}
\label{sec:literature}

There is a large literature concerning polarization, faction formation, and closely related topics across many disciplines.  Attempting to review or summarize all of it would take us too far afield.  Instead, in the present section and the next, we seek to highlight some proposed explanations of epistemic factionalization that we wish to contrast with the mechanism in the model we analyze.

One prominent view is that, because beliefs often polarize along political party lines, the primary explanation for epistemic factionalization is differences of ideology across political parties.  Ideologies, in this sense, are deeply held values or shared metaphors; the idea is that people steeped in different ideologies will be more or less open to accepting various facts and beliefs about the world.  For example, cognitive scientist George Lakoff, in his influential book \emph{Moral Politics}, argues that conservatives and progressives in the United States tend to follow different guiding metaphors or worldviews in understanding politics and key issues \citep{lakoff1996moral}.  On his view, conservatives adhere to a `strict father' model, where the government plays the role of a dominant figure responsible for disciplining wayward citizens.  Progressives, by contrast, adhere to a `nurturing parent' model, where the goal of the government is to keep inherently good citizens protected from harm and corruption.  Disagreement over particular issues results from differences in these worldviews.  As he writes in the 2010 reprint of his book, ``the role of government, social programs, taxation, education, the environment, energy, gun control, abortion, the death penalty, and so on ... are ultimately not different issues, but manifestations of a single issue: strictness versus nurturance" \citep[x]{lakoff1996moral}.

One can think of Lakoff's account as an example of a `common cause' explanation of observed correlation across different beliefs.  Other explanations have a similar structure, but differ over the details of the cause.  For instance, there is a literature in cognitive science arguing that personality traits related to authoritarianism, uncertainty avoidance, and/or social dominance help determine an individual's position on gun rights, taxation, and other policy issues.\footnote{\citet{adorno1950authoritarian} give a very influential treatment of authoritarian personalities.  See \citet{jost2003political} for a good overview of this literature.}  On this approach, it is underlying personality elements that determine why individuals hold the viewpoints they do, and thus provide the common cause of why their beliefs tend to correlate. At the same time, these personality elements tend to determine political party membership, which explains polarization along party lines.

These sorts of common cause accounts have clear explanatory power, and may well contribute to the observed phenomenon.  But as we noted in the Introduction, one also sees correlation in beliefs about seemingly unrelated matters of fact.\footnote{\citet{bramson2017understanding}, in a broader theoretical discussion of polarization, call this `belief convergence'.}  In some cases, there is little reason to think that any particular common cause explains all such examples of correlated beliefs.  This paper will explore an alternative explanation for such patterns of belief polarization.  Rather than supposing there are underlying ideologies or underlying psychologies that explain polarized bundles of beliefs, we focus on the role that social connections play in belief formation.  In doing so, we draw on social epistemology---the study of knowledge-making as it is done by groups of (fallible) humans.\footnote{In a related point, \citet{dellaposta2015liberals} argue that we should not try to explain things like liberal preferences for lattes via appeal to some deep ideological pattern.  As they point out, correlations between latte drinking and liberal politics can emerge as a social phenomenon.}

 In particular, we focus on the role of social trust in epistemic contexts.  Humans tend to trust one another differentially, depending on various factors.  This sort of social trust is deeply important in determining uptake of evidence.\footnote{For instance, a recent study showed that individuals trust articles on Facebook based on their trust in the person who shared them much more than the original source \citep{whosharedit}.}  One such factor is shared belief.  In general, people tend to be more trusting of those who share beliefs with them than of those who do not.  \citet{kitcher1995advancement}, for example, points out that scientists often calibrate their trust in other scientists by comparing others' opinions and beliefs with their own.  In his influential book, \citet{rogers2010diffusion} makes this point in a general way, ``the transfer of ideas occurs most frequently between individuals...who are similar in certain attributes such as beliefs, education, social status, and the like" (18).\footnote{In recent work \citet{marks2019epistemic} found that individuals were less likely to trust individuals with successful track records on an academic task if they believed these individuals did not share their political beliefs.}

In certain ways, calibrating trust to shared beliefs is epistemically well-motivated.  In the case of scientific beliefs, not every source is equally reliable: some scientists are better than others.  Making judgments about reliability of this sort is an essential part of science.  But epistemic reliability is not directly observable, and so scientists must depend on heuristics to ground trust \citep{goldman2001experts}.  One such heuristic is to trust others on the basis of their past epistemic successes.  And for anyone who trusts their own judgment, it makes sense to evaluate other scientists by whether they have come to the same conclusions, since this is evidence of past epistemic success by the lights of whomever is doing the evaluating \citep{o2018scientific}.  But, as we will see, a large modeling literature has shown that grounding trust in shared belief dependably leads to polarization.

\section{Modeling Polarization and Factionalization}
\label{sec:modeling}

Models of polarization begin with the following puzzle: it has been empirically shown that discussion in human groups tends to lead to greater similarity of beliefs.\footnote{See, for example, \citep{festinger1950social} who provide early evidence for this claim.}  Why, then, do we sometimes see stable, interacting groups of individuals who do not share beliefs?  Or whose beliefs grow farther apart over time?  To explain this phenomenon, models of polarization almost universally adopt some version of the following assumption: similarity of belief determines the strength of social influence.  In other words, individuals who share beliefs will also tend to affect each others' beliefs, while those with different views will not.  In this way, models can capture both the fact that individuals tend towards consensus as they engage in discussion, and the fact that this engagement does not always lead to consensus.

There are different ways to instantiate this assumption, and we will not discuss every model that has done so.\footnote{For more complete literature reviews, see \citet{bramson2017understanding} or \citet{o2018scientific}.}  Instead we will give a few relevant examples.  \citet{hegselmann2002opinion} introduce a model of opinion dynamics where individuals start with beliefs between 0 and 1.  In each round, individuals average their beliefs with group members and adopt the average as their new belief.  As they show, if individuals are only influenced by those in some neighborhood of their own beliefs, the group can splinter into subgroups that hold stable, differing opinions.\footnote{\citet{deffuant2002can,deffuant2006comparing} also use this modeling paradigm to explore polarization.}

Most models of this sort focus on opinions, i.e., they do not consider agents who can test the world and gather evidence.  For this reason, they are arguably less applicable to beliefs about scientific matters of fact than to other sorts of beliefs.   \citet{olsson2013bayesian} and \citet{o2018scientific} analyze models of polarization in which agents can explicitly test the world.\footnote{These models differ in that \citet{olsson2013bayesian} focuses on individuals who share statements of belief, while \citet{o2018scientific}, in an attempt to more closely model scientific communities, consider a model where agents share evidence. In addition, \citet{Singer2017} consider a polarization model where agents share `reasons' for a belief, in the form of positive and negative weights, which might be interpreted as evidence from the world.}  As these authors show, the sort of tendency described---trusting only those who share similar beliefs---can dependably lead to polarization.

In the present paper, however, we are interested in more than just polarization.  We wish to study how multiple, unrelated, polarized beliefs can come to correlate with each other.  Most models of polarization do not look at multiple beliefs.  There is, however, a distinct literature on cultural diffusion that captures some aspects of faction formation.  \citet{axelrod1997dissemination} presents a model in which multiple cultural attributes determine whether or not two cultures will become more similar when they meet.  His agents, which represent small groups such as interacting villages, consist in a list of numbers representing cultural attributes, each of which can take several different values.  Agents are arranged in a grid.  Pairs of neighbors are randomly selected, and adopt cultural attributes from each other with a likelihood given by the proportion of shared attributes.  As he shows, over time cultural similarity tends to increase, but if two cultures do not share any attributes they will remain stably different.  One of his conclusions is that, ``when cultural traits are highly correlated in geographic regions, one should not assume that there is some natural way in which those particular traits go together" (220).  In other words, grounding cultural adoption in multiple attributes leads to correlation among those attributes.

Subsequently, a literature has emerged looking at variations of this model, largely in the interest of exploring the emergence and stability of cultural diversity.  For instance, does a small amount of cultural drift---random adoption of new variants---lead to homogeneity \citep{klemm2003global,centola2007homophily}?  How does information transfer via media influence cultural adoption \citep{shibanai2001effects}?  How do logical connections between beliefs influence all this \citep{friedkin2016network}? In addition, \citet{dellaposta2015liberals} give a network model where individuals have a number of ``lifestyle traits".  Trait similarity influences interaction, and interaction tends to lead to similarity.  They find the emergence of correlation among these traits.  The important take-away is that in models where multiple attributes ground influence, we see correlation between those attributes.

While these models can incorporate opinions or beliefs as cultural attributes to some degree, they do not capture the idea that some beliefs concern matters of fact.  In particular, there is no sense in which agents are able to gather or share evidence about their beliefs; and no sense in which holding false beliefs may reduce agent payoffs.   Here we investigate a model where, unlike the Axelrod and similar models, agents attempt to achieve true beliefs, and in doing so are able to gather and share evidence.  Unlike previous models of polarization, they have more than one belief and ground their trust in evidence over these multiple arenas.  As we will see, we can investigate how polarized beliefs might come to correlate even among agents who use evidence to attempt to learn the truth.

\section{The Model}
\label{sec:model}

Our model begins with the network epistemology framework developed by \citet{venkatesh1998learning}.  This model was brought into philosophy of science by \citet{zollman2007communication} to represent discovery within a scientific community.  Since then it has been used to ask questions such as: what is the optimal communication structure for the spread of scientific belief?  How does conservatism influence scientific progress?  And how might industrial propagandists influence scientific communities?\footnote{For more on the use of this framework in philosophy of science see \citet{zollman2010epistemic,mayo2011independence,kummerfeld2015conservatism,holman2015problem,rosenstock2017epistemic,Weatherall+etal,OConnor+WeatherallConformity,OConnor+WeatherallBook, borg2017examining, freyrobustness, freywhatis, holman2017experimentation}. \citet{zollman2013network} provides a review of the literature up to 2013.}

The original model has two key elements: a decision problem and a network.  In more detail, there is a collection of some number of agents, $N$, arranged on some network.  Each of these agents faces a decision problem where they decide which of two actions, $A$ or $B$, to take.  For example, maybe they are doctors who will prescribe one of two pills for morning sickness, and they need to pick which.  Each of these actions yields the same payoff on success, but with different success rates, $p_A$ and $p_B$, respectively.  For the version of the model we look at here, we assume that the success rate of $A$ is well established to be $p_A = .5$.  The problem is that actors are unsure whether the success rate of $B$ is better, $p_B = .5+ \epsilon$, or worse, $p_B = .5-\epsilon$.  In reality, $B$ is better.  (For mnemonic purposes, we can say that $A$ is for `All-right' and $B$ for `Better'.)   This decision problem is sometimes called a `two-armed bandit problem'.  A one-armed bandit is a slot machine, and so this problem is one in which an individual would prefer to pull the more profitable of two arms on a slot machine, but where they are unsure which, in fact, pays out more.

We suppose that each agent has some credence, represented by a number in the interval $(0,1)$, that $B$ is the better arm.  Initially, these credences are drawn from a uniform distribution over the interval of possible credences.  In each round of simulation, agents perform the action that they think is better some number of times, $n$.  This might represent the doctor prescribing the pill that they believe is more likely to work, or the gambler pulling the arm they think pays out more, $n$ times.  Then, based on the results of the action, each agent updates their beliefs concerning action $B$ using strict conditionalization (i.e., Bayes' rule).  For example, an agent with credence .54 will perform action $B$ $n$ times, resulting in some number $k$ of successes, and then update their beliefs accordingly.  An agent with credence .38 will pull arm $A$, and since this arm is already well understood, they will not change their beliefs at all based on the results.

Agents in this model do not update only on their own results; they also update on the results of their neighbors in the network.  The assumption is that individuals share their evidence with those they are connected to.  In our models, we will use a trivial network structure---the complete network---where agents see evidence from all others.  As will become clear, we do this to emphasize how polarization can emerge despite the fact that all agents have the potential to engage with each other.  Note, though, that for this reason, it might be just as easy to think of this as a model without a network.  Or alternatively, as will become clearer, one could think of it as a model with an evolving network where the strength of links tracks trust in other agents.

In simulations of the model as described so far, with all agents treating all evidence to which they are exposed identically and using Bayes rule to update their beliefs, groups of agents always end up in a state of consensus.  That is, either all agents come to have a very high credence that action $B$ is, in fact, better, or else, all agents come to falsely believe action $A$ is better, and because they stop testing $B$ they never adopt more accurate credences \citep{zollman2007communication, zollman2010epistemic, zollman2013network, rosenstock2017epistemic}.  In other words, because of the mutual influence agents have on each other, stable polarization does not emerge.\footnote{\label{fn:stability} Note that false consensus is always stable in this model, because no agents test action $B$.  True consensus, on the other hand, is not strictly stable, as stochastic effects, in the form of sufficiently long strings of spurious results, can always push agents from true consensus to false consensus.  However, the probability of this occurring goes to zero as the beliefs of the agents approach 1.  We remark that disagreement occurs on the way to consensus, but this does not capture the phenomenon of polarization, in which disagreement is, at least approximately, stable.}

We modify this basic model in two ways.  The first modification follows \citet{o2018scientific}, who consider a variation of this model in which agents do not treat all evidence in the same fashion.  Instead, they treat evidence from their neighbors as \emph{uncertain}, reflecting that they do not entirely trust other agents, and then update their beliefs using \emph{Jeffrey conditionalization}, an alternative to strict conditionalization introduced by philosopher Richard Jeffrey \citep{Jeffrey} to capture rational belief revision in the presence of uncertain evidence.  In particular, O'Connor and Weatherall introduce the assumption that agents become more uncertain about evidence the further their beliefs are from those of the neighbor who produced that evidence.  Two agents who share similar beliefs will treat each others' evidence as trustworthy, and update strongly on it.  Two agents whose beliefs differ greatly will lose trust in each other, and update only weakly, or not at all, or even in the other direction, on the evidence provided.  This, again, is the sort of dependence between shared belief and social influence that allows for polarization.

Let us now look in more detail at how updating works in the model.  First, let us consider Jeffrey's rule.  Suppose that an agent, Nicolas Cage, is presenting evidence to Liam about hypothesis $H$, but Liam does not fully trust it.  He has credence $P_f(E) \leq 1$ that his evidence ($E$) really obtained.  Then he will update jis belief in $H$ using the following formula:
\[P_f(H) = P_i(H|E)\cdot P_f(E) + P_i(H|\sim E) \cdot P_f(\sim E)\]
This equation says that his new credence in the hypothesis ($P_f(H)$) is equal to the probability of the hypothesis under strict Bayesian updating on the evidence ($P_i(H|E)$) times his credence the evidence is real ($P_f(E)$), plus the probability of the hypothesis under Bayesian updating if the evidence did not obtain ($P_i(H|\sim E)$) times his credence that the evidence did not obtain ($P_f(\sim E)$).  In other words, his uses Bayesian updating under two weighted possibilities---that the evidence obtained and that it did not.

The key question now is: how do we determine his level of mistrust in Nicolas?  How does an agent calculate $P_f(E)$ for some other agent?  In \citet{o2018scientific}, agents develop beliefs about one issue, and $P_f(E)$ is a decreasing function of the difference, $d$ between their credences, using the following formula:\footnote{\citet{o2018scientific} consider several functions: a linear function, as below, as well as logistic and exponential functions, and found that the results were stable across these modeling choices.}
\begin{equation}P_f(E)(d) = \max(\{1-d\cdot m \cdot (1-P_i(E)),0\}).\label{antiUpdate}\end{equation}
Here $P_i(E)$ is the agent's prior probability that someone performing action $B$ would yield outcome $E$, while $P_f(E)$ is the agent's belief that $E$ was, in fact, obtained, given that this evidence was reported to them by an agent whose beliefs are distance $d$ from their own.

\begin{figure}
\centering
\includegraphics[width=.9\textwidth]{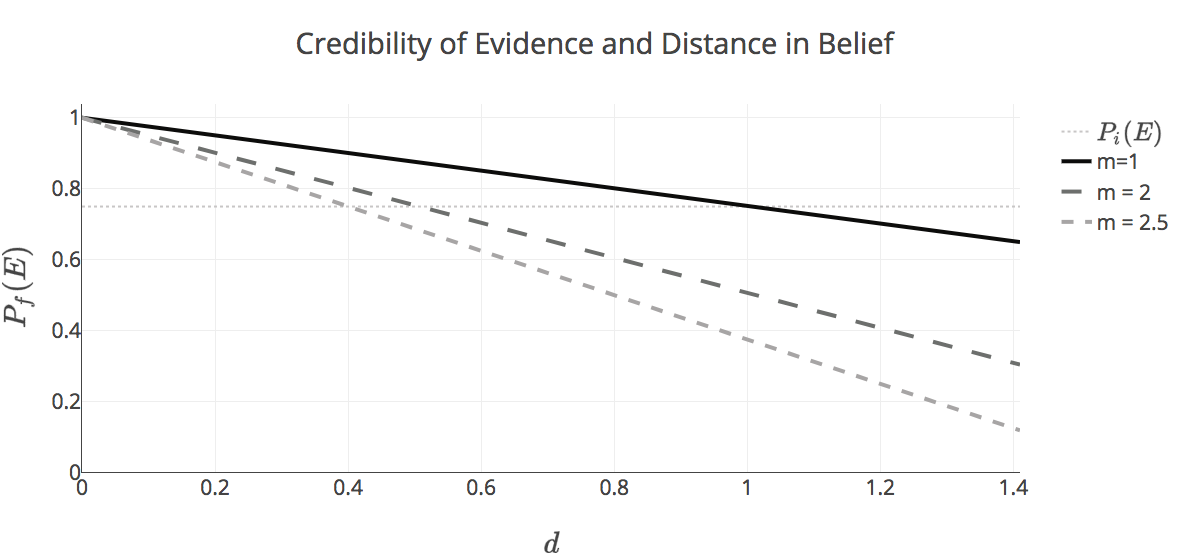}
\caption{Credence in evidence for three values of $m$, the multiplier that determines how quickly agent mistrust increases in distance between beliefs.  The prior probability of the evidence is $P_i(E) = .75$.\label{fig:creddistance}}
\end{figure}

Observe that, respecting the fact that one cannot assign negative probabilities to events, $P_f(E)$ is bounded from below by $0$.  Otherwise, the function takes the value 1 when distance is 0 (i.e., agents with identical beliefs treat one another's evidence as certain) and decreases linearly with distance $d$, at a rate determined by a parameter $m$, which characterizes the strength of `mistrust'.\footnote{Observe that this means that all agents privilege their own evidence, necessarily treating it as certain.}  This result is scaled by the distance between 1 and the prior probability of the evidence, which captures the idea that if the evidence is very probable on an agent's prior beliefs, it does not become highly unlikely just because someone shared it.  Figure \ref{fig:creddistance} shows what this function looks like for $m=1,2,2.5$. (In this figure we suppose the updating agent has prior probability $P_i(E)=.75$ that the evidence would obtain.)  Notice that the higher $m$ is, the more quickly uncertainty increases with distance.

Also notice that on this model, it is possible for an agent's credence that the evidence occurred to be \emph{less} than it would have been before another agent reported their beliefs.  This behavior, which we call `anti-updating' in what follows, is connected to a psychological phenomenon known as the backfire effect \citep{cook2016rational}. \citet{o2018scientific} also consider models where agents simply ignore evidence once their trust function reaches their prior probability of evidence.  In figure \ref{fig:creddistance}, this would involve bounding $P_f(E)$ from below at .75.  This alternative, which we will call `no anti-updating,' assumes there is no backfire effect.  In generating results, we consider both possibilities.

The second alteration to the basic model, which we study in the present paper for the first time, is to consider what happens when this sort of epistemic trust is grounded in multiple beliefs.  In other words, we consider agents who assess whether to trust evidence from some source by looking at similarity of belief in multiple areas.  This might correspond to a vaccine skeptic who places more trust in other vaccine skeptics when assessing the efficacy of homeopathy than they do in someone who believes vaccines are safe.

More precisely, we suppose agents are presented with two, unrelated decision problems, both of which are two armed bandit problems as described above.  Each agent's epistemic state is represented by two distinct credences, $P(B_1)$ and $P(B_2)$, both valued in the interval $(0,1)$.  The first credence, $P(B_1)$, represents the agent's belief about the first action; the second credence, $P(B_2)$, represents that agent's belief about the second action.  Once again agents begin with random beliefs, and then in each round, perform whichever actions they believe to yield the best outcome $n$ times, and then update their beliefs in light of their result and the results of their neighbors.  Now, however, they perform both actions in each round, and apply Jeffrey's rule to update their beliefs in light of the evidence to which they have been exposed.

\begin{figure}
\centering
\includegraphics[width=.4\textwidth]{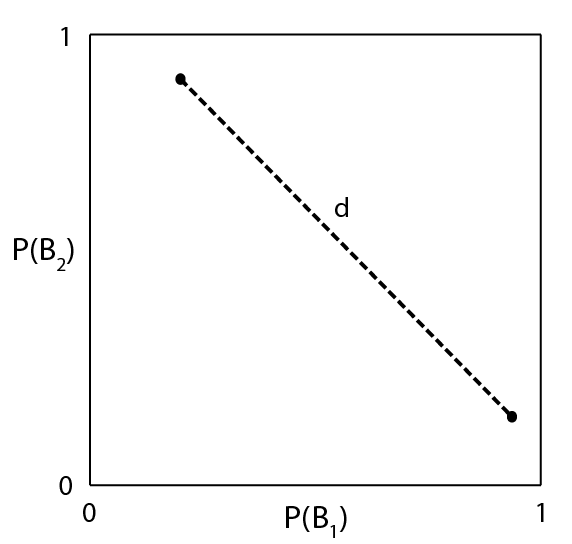}
\caption{The space of beliefs on two topics for two agents.  Each dot represents one agent's beliefs.  The line between represents the distance, $d$, between these agents in belief.\label{fig:distance}}
\end{figure}

Following \citet{o2018scientific}, we use Eq. \eqref{antiUpdate} (or the modified version, without anti-updating) to calculate the credence agents assign to evidence produced by their neighhors, except that now we suppose trust is grounded in similarity of belief on both issues.  We do this by taking $d$ to be the Euclidean distance between the two agents in a two dimensional `belief space' with coordinates $(P(B_1),P(B_2))$.\footnote{To be explicit: The distance between beliefs, $d$, is the Euclidean distance between them, $\sqrt{(P_1(B_1)-P_2(B_1))^2 + (P_1(B_2)-P_2(B_2))^2}$, where indices on $P$ reflect the two different agents.}  Figure \ref{fig:distance} shows what this looks like. Each dot represents one agent's beliefs.  The minimum distance between two agents is again, $d=0$, which obtains when they agree precisely on both beliefs.  The maximum distance is $d=1.41$, which occurs when agents are on two opposite corners.   (For instance, one has credence 1 in $B_1$, credence 0 in $B_2$, and the other has credence 1 in $B_2$ and credence 0 in $B_1$.)

As we will discuss in section \ref{sec:results}, there are a few further variants we consider in what follows.  We look both at situations where agents have already polarized on one belief, and situations where they develop beliefs in more than two arenas at once.  In addition, we consider many parameter settings.  We vary $n$, the number of draws on each arm that agents take every round.  This parameter helps determine the quality of evidence that agents share. We vary $\epsilon$ which determines how much more often the better arm pays off than the worse one.  This corresponds to how much more successful the two better theories are than their baselines.  We vary the size of the group, $N$.  And, most crucially, we vary the strength of $m$, which controls the level of distrust in other agents.  Each combination of parameters was run 1024 times, and reported results are averages over these runs.  We run simulations until all agents reach a (approximately) stable state; the possible outcomes are described in detail in the next section.

\section{Results}
\label{sec:results}

 There are several stable outcomes that can arise in these models.\footnote{Or, to be more precise, approximately stable; recall the considerations in note \ref{fn:stability}.}   First, sometimes all agents in a simulation will come to consensus on both decision problems.  This consensus can be to the true belief in both cases (defined as all agents having credence $>=.99$), the false belief in both cases (defined by all agents having credence $<.5$), or the true belief for one case and the false for the other.  Second, sometimes agents polarize over one problem, but not the other.  And last, sometimes agents polarize over both problems.  Polarization, in the context of this model, occurs when some agents in a simulation develop high credences ($>.99$) in the better action, $B$, and others hold low credences ($<.01$ with anti-updating, or $<.5$ without anti-updating) in action $B$.  To count as a `polarized' outcome, we require that the polarization be stable, which occurs when the multiplier, $m$, is such that the beliefs of the $A$ agents are outside the range where they update on the evidence coming from the $B$ agents.\footnote{Note that if $m$ is such that agents always update on the evidence from all others, at least a little bit, transient polarization is possible, but agents eventually reach consensus.  Whenever $m \leq 1/\sqrt{2}$ in our two-problem models, all agents will have some influence on all others.}   For anti-updating models, the $A$ agents' beliefs will be driven down close to zero as they anti-update on evidence coming from $B$ agents.  For no-anti-updating models, $A$ agents will have a variety of stable, low credences, depending on $m$.

When agents polarize on both problems, there can be different levels of correlation between the two beliefs.  To measure correlation, we see whether, at the end of simulation, each agent has settled on the true or false belief for each problem.  From this, we construct two $N$ dimensional vectors, $x$ and $y$, whose entries consist of $0$ and $1$, for false and true beliefs, respectively.  We then calculate the (sample) Pearson correlation coefficient $r$ for the two vectors:
\[
 r={\frac {\sum _{i=1}^{N}(x_{i}-{\bar {x}})(y_{i}-{\bar {y}})}{{\sqrt {\sum _{i=1}^{N}(x_{i}-{\bar {x}})^{2}}}{\sqrt {\sum _{i=1}^{N}(y_{i}-{\bar {y}})^{2}}}}},
\]
where $\bar{x}$ and $\bar{y}$ are the means of vectors $x$ and $y$ respectively, and subscript index $i$ indicates the $i$th component of each vector.  This yields a measure of the correlation between the two sets of beliefs of the agents, valued in the interval $[-1,1]$, where $1$ represents perfect correlation and $-1$ perfect anti-correlation.  We then take the absolute value of $r$, and average this number over simulation runs.

Why take the absolute value?  Consider a situation in which three agents believe that both $B_1$ and $B_2$ are the better actions (i.e., hold true beliefs about both problems), and three other agents believe both $B_1$ and $B_2$ are the worse actions (i.e., hold false beliefs about both actions).  In this case, there would be perfect correlation ($r = 1$).  On the other hand, if three agents believe $B_1$ is better and $B_2$ worse, and the other three believe $B_1$ is worse and $B_2$ better, we would find there is perfect anti-correlation ($r = -1$).  Notice, however, that both of these scenarios correspond to the sort of phenomenon we are trying to capture: endogenous factions have formed where bundles of otherwise unrelated beliefs are shared.  It is for this reason that we focus on the absolute value of correlation between beliefs at the end of simulation.  This will be a number between 0 (no correlation in the relevant sense) and 1 (perfect correlation).

We compare the average level of correlation for each set of parameters with what we would expect if trust were not grounded in multiple beliefs---that is, what we would expect if the agents developed beliefs about each problem separately, using only distance in the one-dimensional belief space associated with each individual problem to determine their uncertainty about evidence.\footnote{To determine these baseline cases, we ran simulations with identical parameters, but in which the model dynamics was altered so that the agents treated each of the problems separately, and then determined the average absolute value of $r$ in these simulations.}  This correlation is generally greater than 0, both because of random chance, which pushes the absolute value of correlation above 0, and because in these models, agents are drawn towards true beliefs, which means that even when agents polarize, more than half of them end up at the true belief.  (Details of the parameter values for the model determine how likely it is that agents develop true beliefs \citep{zollman2007communication, zollman2010epistemic, rosenstock2017epistemic, o2018scientific}.)  Furthermore, the size of the group determines how much correlation one should expect from stochasticity.  Thus the `baseline' correlation is different for each set of parameter values of the model.

\subsection{Pre-existing Polarization}\label{sec:prePol}

To get a handle on how the model works, we begin by discussing what is arguably the simplest treatment we will consider. Suppose that a community is already polarized with respect to one problem.  This might correspond to a scenario, for example, where individuals are already polarized with respect to the dangers of vaccines, and begin to develop beliefs about a new issue---the dangers of genetically modified crops, for instance.  In particular, we suppose that at the beginning of simulation, half of agents have $P(B_1) = 1$ and the other have $P(B_1) = 0$.  In other words, half of agents hold the true belief regarding problem one with certainty, and the other are certain it is false.  We then randomly initialize credences about the second problem, $P(B_2)$, and let agents update as described in section \ref{sec:model}.  Notice that in this model, agents in the two camps will always have a distance in belief of $d \geq 1$.  The question is: will polarization over the new problem reflect the same fault lines as those of the established beliefs?

\begin{figure}
    \centering
    \includegraphics[width=0.8\textwidth]{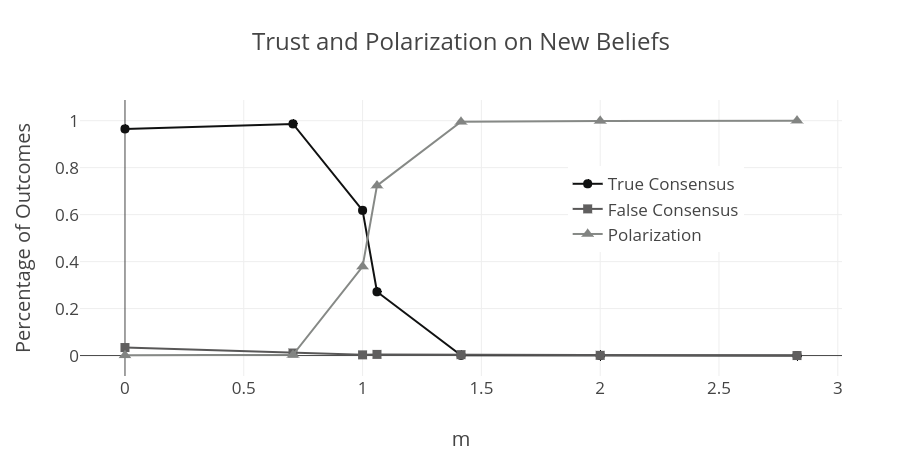}
    \caption{Simulation outcomes as $m$ increases in models where agents start with one polarized belief.  These results are for $N = 20$, $n = 50$, $\epsilon = .01$, and anti-updating.}
    \label{fig:polarprepolar}
\end{figure}

In this model, it is stipulated that agents are polarized with respect to the first problem, so there are now only three kinds of outcomes.  Either agents all reach true consensus on the second problem, false consensus on the second problem, or they polarize with respect to this problem as well.  Regardless of other parameter settings, polarization about the second problem becomes increasingly likely as $m$ increases.  This is unsurprising, since $m$ is the key parameter establishing how quickly agents stop trusting each other.  Furthermore, when agents anti-update---actually mistrust evidence more if it is shared with someone who has very different beliefs---polarization is always more likely than if agents do not.  Figure \ref{fig:polarprepolar} shows proportions of simulation outcomes that ended with true, false, or polarized outcomes for one set of parameters.\footnote{In general, we randomly choose parameter values to illustrate each point we make; all of the trends we consider are stable across parameter choices.} As is evident, as $m$ increases there is a dramatic uptick in polarized beliefs.

\begin{figure}
    \centering
    \includegraphics[width=0.8\textwidth]{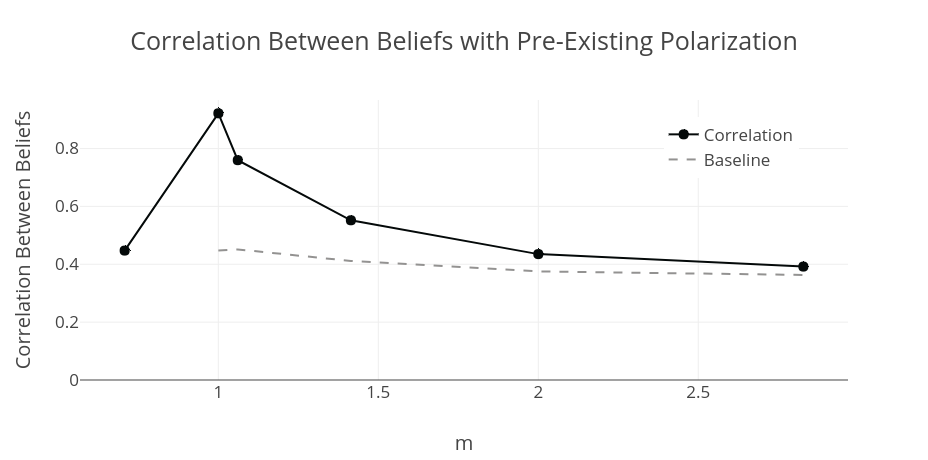}
    \caption{Average correlation between beliefs as a function of $m$ for models where agents start with one polarized belief.  These results are for $N = 6$, $n = 20$, $\epsilon = .1$, and anti-updating.}
    \label{fig:corprepolar}
\end{figure}

When agents depend on two beliefs to ground trust, and when they are already polarized on one of these, correlation emerges between the beliefs.  We look, in particular, at the level of correlation that emerges when agents reach a polarized outcome.  (Otherwise correlation is zero, since agents entirely agree about the second problem.)   Figure \ref{fig:corprepolar} shows average correlation for simulations with $N = 6$, $n = 20$, $\epsilon = .1$, and anti-updating as $m$ varies.\footnote{Notice that because levels of polarization vary across values, the results in this figure are averaged over different numbers of simulations for each data point.}  The level of correlation, as mentioned, is compared to a baseline level of what would be expected if beliefs were developed independently.  As is evident, for all values of $m$ there is more correlation than would be randomly expected.  Although the level varies with different parameter values, this was true across our data set.\footnote{We remark that comparing values of $m$ between the baseline and the full model, is somewhat subtle, because the minimum value of $m$ at which agents can possibly polarize differs in the two models.  We have not attempted to correct $m$ in these figures; one can imagine, if one likes, translating the dotted line to the left so that the first values at which non-zero correlation occurs coincide.}

There is an observable trend in this figure, which is that when $m$ becomes larger than 1, correlation decreases between the two polarized beliefs.  This trend is also visible for different parameter settings.  This might be surprising, since $m$ is the parameter that controls the amount of social trust across beliefs.  What is happening in these cases is that as $m$ becomes large, agents in this model break into four polarized camps reflecting the extremes for each belief: true for $B_1$ and false for $B_2$, true on both, false on both, or false for $B_1$ and true for $B_2$. This more dramatic polarization makes the Pearson correlation coefficient a less effective measure of the correlation between beliefs on the two problems. But note that in these cases we still have four epistemic factions forming.  In each members hold a bundle of beliefs and now do not trust others who have any other such bundle.

\begin{figure}
    \centering
    \includegraphics[width=0.8\textwidth]{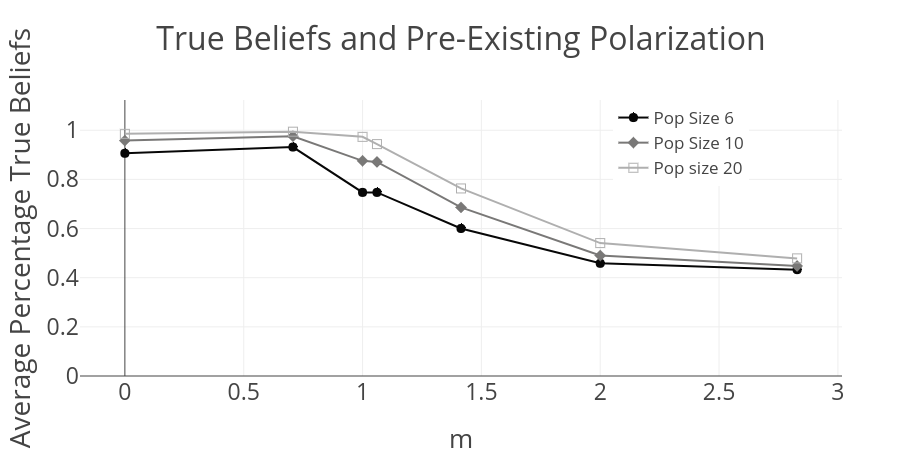}
    \caption{Average number of true beliefs as a function of $m$ for models where agents start with one polarized belief.  These results are for  $n = 20$, $\epsilon = .05$, and no anti-updating.}
    \label{fig:truthprepolar}
\end{figure}

The last result we wish to pull out is that, in general, as $m$ increases, agents become increasingly unlikely to reach true beliefs.\footnote{Modulo one exception, discussed below.}  Figure \ref{fig:truthprepolar} shows the average percentage of true beliefs held, on belief two, as $m$ increases.  This is for different population sizes with $n = 20$, $\epsilon = .05$, and no anti-updating.  (Results were highly similar for the anti-updating version of the model.  They were also highly stable over different parameter values.)  As is evident, as $m$ increases, true belief decreases sharply.\footnote{There is one exception, which occurs when $m$ goes from 0 to $1/\sqrt{2}$.  The average percentage of true beliefs increases slightly.  This seems to be because when $m=1/\sqrt{2}$, all actors will eventually reach consensus, but $m\neq 0$ slows learning.  As \citet{zollman2007communication, zollman2010epistemic} has shown, there is sometimes a benefit in this sort of epistemic network model to slow learning processes where actors do not too quickly lock into possibly false beliefs.  This likely explains the small increase in true beliefs at this value.}

We emphasize that this phenomenon, of increasing $m$ reducing the number of true beliefs, occurs not because the number of false consensus outcomes increases markedly with $m$, but rather because in stably polarized outcomes, which become more common as $m$ increases, the fraction of the population holding the false beliefs increases with $m$, approaching $.5$ on average.  By some measures of polarization, this means that larger $m$ tends to mean more polarization, in the sense not only that the population includes agents whose beliefs have diverged, but also that the relative sizes of the two camps tend to approach one another \citep[cf.][ for a discussion of such measures]{bramson2017understanding}.  Conversely, for smaller values of $m$, polarized outcomes tend, on average, to have a majority of agents holding the true beliefs, and a minority stably holding the false beliefs.  One might argue that this is a situation that is \emph{less} polarized, even though it is stable and agents' beliefs have diverged.  While we acknowledge that there are interesting measures of polarization for which this is true, we would point to real-world examples, such as vaccine skepticism, where beliefs appear to be approximately polarized, only a relatively small fraction of the population holds the false belief, and yet because of background facts about the case, that small fraction of people holding the false belief can create serious public health risks, as evidence that stable polarization with highly unequal camp size can nonetheless be socially significant.

Why do false beliefs increase with $m$?  In these models, it is assumed that all agents are drawing unbiased data from a distribution.  In other words, all agents have good evidence to share.  Mistrust leads individuals to ignore the good evidence in their community, or even to update in the opposite direction.  Furthermore, in this model those with accurate beliefs are also those gathering evidence about the better action.  This means that the individuals with inaccurate beliefs are ignoring exactly those community members whose evidence would lead to better beliefs.

We conclude this subsection by observing that the version of the model discussed here has another interpretation.  Since the agents begin by being dogmatic about action 1, their credences do not change in light of evidence.  This means that one might as well suppose that these values do not represent credences at all. Instead, one might take them to represent some other `social identity', such as political party or race, along which the agents differ, and which factors into the agents' trust of one another.  On this interpretation, the results we have described show how agents who differ in social identity, and who systematically mistrust evidence from agents with a different identity, can evolve to hold polarized beliefs about other topics in a way that correlates with their identity.  The important thing to emphasize is that there is no connection between the `content' of belief about action 2 and the agents' identity.  Thus, though we can see how correlation between beliefs and something such as political party can emerge in our model, there is no `common cause' explanation in the sense discussed in section \ref{sec:literature}, because the social identity of the agents does not explain why those agents came to hold one belief rather than the other.  (That is, it is not that there is a `single issue' with many manifestations, or a `personality profile' with many consequences.)  Agents polarize along pre-determined lines, but which group ends up holding which belief is stochastic and varies from simulation run to simulation run.

\subsection{The Co-Evolution of Polarized Beliefs and Endogenous Factionalization}

In the last subsection we considered a model where agents started off polarized in their beliefs about one problem, and used this to ground trust concerning a second problem.  What we found was that the initial polarization influenced agent outcomes, leading to correlation between the two beliefs.  Now let us suppose that agents develop beliefs about two topics at once.  At the start of simulation, agents are initialized with random credences about both problems.  They can thus end up at any of the five outcomes described at the beginning of this section.

\begin{figure}
    \centering
    \includegraphics[width=0.8\textwidth]{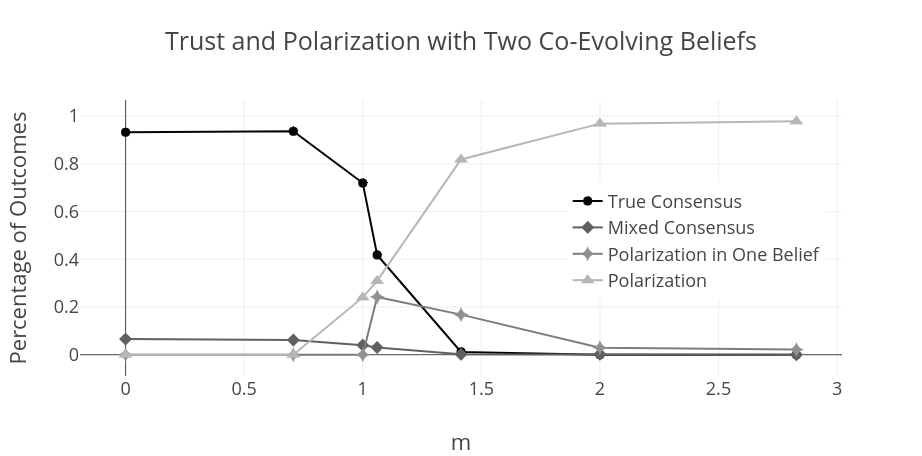}
    \caption{Outcomes as a function of $m$ for models where agents develop two beliefs.  These results are for $N = 10$, $n = 5$, $\epsilon = .1$, and no anti-updating.}
    \label{fig:polar2B}
\end{figure}

We find that as mistrust increases, polarization in both beliefs increases.  (See figure \ref{fig:polar2B}.)  When $m$ is small, the most likely outcome by far is consensus on true beliefs, though some simulation runs end up with true consensus on one belief, and false on the other---labeled `Mixed Consensus'.  (For other parameter values, false consensus on both beliefs will also occur.  It just happens to be vanishingly unlikely for the parameters in this figure.)  As $m$ increases, the two other sorts of outcomes become more likely.  First, it is sometimes the case that agents settle on consensus for one belief, but are polarized on the other.  And as $m$ increases further it becomes practically guaranteed that agents will be polarized on both beliefs.

Given some value of $m$, polarization is more prevalent in these models than in ones where actors ground trust on each separate belief.  This is essentially because actors have greater opportunity for divergence of beliefs when they look at multiple problems at once, because the maximum distance between agents increases.  These divergent beliefs lead to greater mistrust and more polarization.

\begin{figure}
    \centering
    \includegraphics[width=0.8\textwidth]{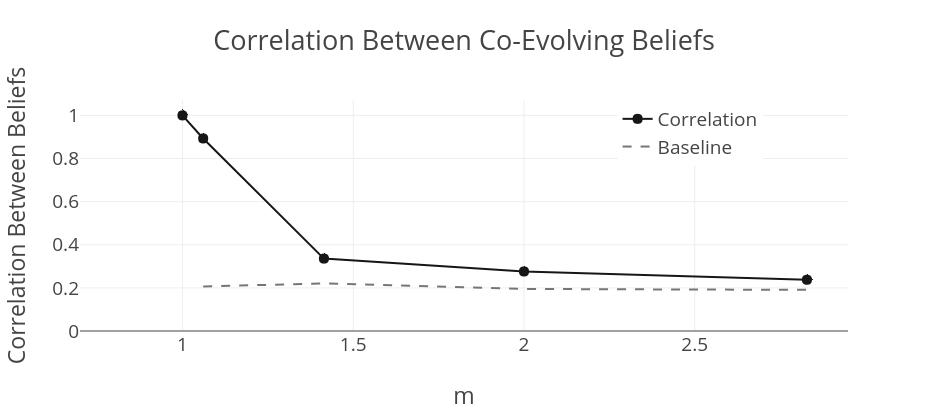}
    \caption{Correlation as a function of $m$ for models where agents develop two beliefs.  These results are for $N = 20$, $n = 50$, $\epsilon = .05$, and no anti-updating.}
    \label{fig:cor2B}
\end{figure}

Once again, we find that beliefs on these two problems correlate when both are used to ground trust.  And once again, while the level of correlation between beliefs will depend on various parameter values, it is always the case that this correlation is higher than would be otherwise expected.  Now, however, this correlation arises endogenously, in the sense that agents evolve to hold polarized beliefs about both problems, and do so in a highly correlated way.  Figure \ref{fig:cor2B} shows correlation for $N = 20$, $n = 50$, and $\epsilon = .05$, though again the qualitative results were stable across parameter values.

\begin{figure}
    \centering
    \includegraphics[width=0.8\textwidth]{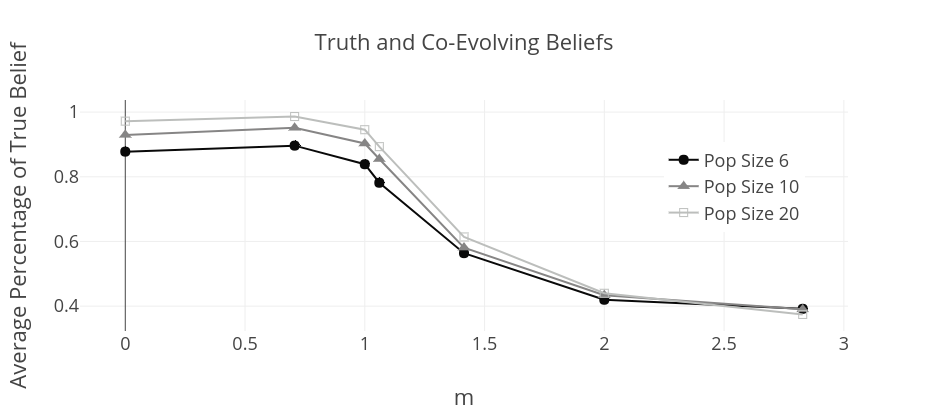}
    \caption{Average true beliefs as a function of $m$ for models where agents develop two beliefs.  These results are for $n = 10$, $\epsilon = .05$, and anti-updating.}
    \label{fig:truth2B}
\end{figure}

We also find, as in the pre-polarized treatment, that increasing the mistrust parameter $m$ decreases the average number of true beliefs at the end of simulation.  Figure \ref{fig:truth2B} shows average true beliefs for different population sizes as $m$ increases.  Once again, it is clear that they drop off as mistrust increases.

The key takeaways, here, are that endogenous epistemic factionalization can occur, in simple models with highly idealized agents, without assuming \emph{any} `common cause' explanation.  Instead, agents align their beliefs with one another because they trust one another differentially, in a way that depends on their distance in beliefs about two different issues.  As simulations progress, agents find themselves listening to the same other agents on both topics, leading to situations where those agents' beliefs evolve together.  Moreover, this dynamic leads to situations in which, on average, the entire group's epistemic situation is worse than if they did not adopt the trust heuristic, insofar as the prevalence of false beliefs increases.  As we discuss below, the fact that a common cause is not necessary to explain epistemic factionalization does not mean that common causes do not exist.  But it does suggest that one is not compelled to adopt \emph{prima facie} procrustean arguments about how apparently different topics are nonetheless connected.  Epistemic factions can emerge as a result of trust dynamics alone.

\subsection{Polarization in Multiple Beliefs}

The results discussed so far have concerned agents solving two problems at once.  We also consider a more complicated version of the model, in which agents develop beliefs about three problems at the same time.  The goal is to see whether the lessons from the last two sections---about polarization, correlation, and truth---hold up when we look at a slightly more complex case.  To be clear, at the beginning of these simulations, agents are initialized with three random credences about three possible topics.  They use $m$, and distance in belief, to determine the strength of updating on the evidence of others in their group, as in Eq. \eqref{antiUpdate}.  This time, however, $d$, is the Euclidean distance between agents' credences in three dimensional belief space.  The largest possible distance is $d= 1.73$, which is achieved when agents have opposite credences on each belief, i.e., when they lie in `opposite corners' of a cube with unit sides.

\begin{figure}
    \centering
    \includegraphics[width=0.8\textwidth]{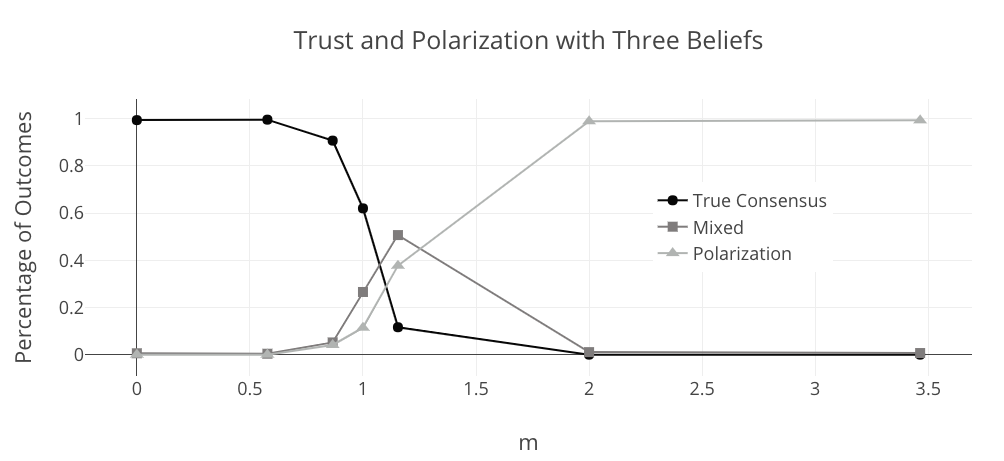}
    \caption{Outcomes as a function of $m$ for models where agents develop three beliefs.  These results are for $N = 10$, $n = 10$, $\epsilon = .2$, and no anti-updating.}
    \label{fig:polar3B}
\end{figure}

In the three belief model, the number of possible stable outcomes is much larger than in the two belief model.  Now we can have outcomes where everybody has true beliefs in each arena, everybody has false beliefs in each arena, there is consensus in each arena, but some are true and some false, there is consensus over some beliefs but polarization on others, or all beliefs are polarized.  Figure \ref{fig:polar3B} shows what outcomes tend to look like as mistrust increases in this model.  For simplicity sake, we show three outcomes---true consensus, polarization in all beliefs, and a mixture of the other outcomes together.  (False belief in all arenas never occurred for this set of parameters.)  Results are for $N=10$, $n=10$, $\epsilon=.2$, and no anti-updating, but are qualitatively similar over other parameter choices.  As is evident from this figure, once again true beliefs are highly likely when $m$ is small.  As $m$ increases the chances of one of these `other' outcomes increases and then decreases; for $m$ sufficiently large, agents essentially always polarize on all beliefs.

\begin{figure}
    \centering
    \includegraphics[width=0.8\textwidth]{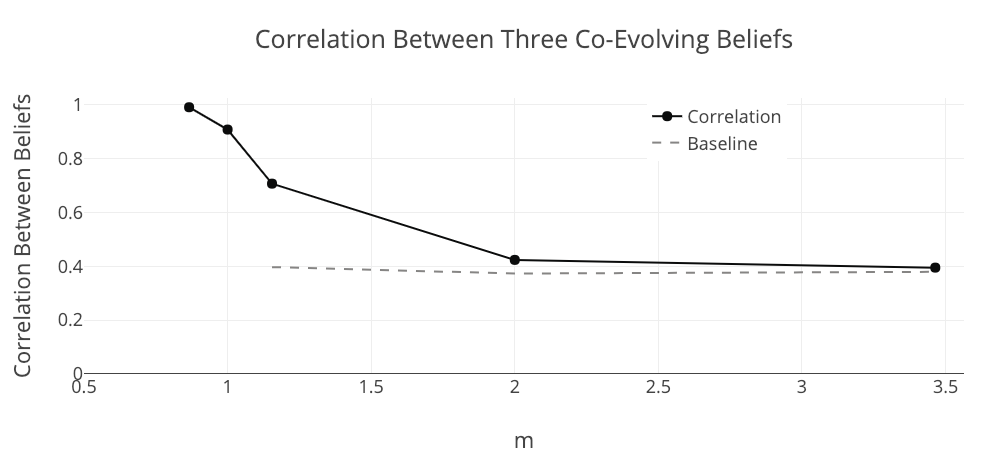}
    \caption{Correlation as a function of $m$ for models where agents develop three beliefs.  These results are for $N = 6$, $n = 50$, $\epsilon = .o1$, and anti-updating.}
    \label{fig:cor3B}
\end{figure}

To study how well beliefs correlate in this model, we calculate correlation coefficients pairwise between beliefs about each problem.  That is, we consider the correlation coefficient between beliefs about problems one and two, one and three, and two and three, and then average the absolute values of these to yield a total level of correlation between beliefs. We again find that when actors ground trust in beliefs about multiple topics, their beliefs ultimately correlate more than would be expected on the baseline analysis.  And once again, this correlation is strongest for intermediate values of $m$, because large $m$ leads to polarization in multiple `directions' in the space of beliefs. Figure \ref{fig:cor3B} shows correlation between beliefs as a function of $m$, for $N = 6$, $n = 50$, $\epsilon = .01$, and anti-updating.  The increased correlation over what would be expected at baseline was generally less dramatic in this model.  This is, in part, because the three belief structure creates more opportunity for polarization along multiple dimensions of belief.  For some levels of $m$, however, the difference is dramatic and significant correlation occurs.

Though we will not show a figure for this, increasing $m$ also decreases the number of true beliefs in this version of the model.  This occurs across our data set.

\section{Conclusion}
\label{sec:conclusion}

We show in this paper that a simple heuristic---to mistrust those who do not share one's beliefs---leads not only to polarization, but to correlation between multiple polarized beliefs even in cases where ample evidence is available.  This may help explain why we so often see beliefs with strange bedfellows, especially in cases where there is profound mistrust of those with different views.  As we have emphasized throughout the paper, we find this result without assuming any kind of common cause: the agents do not share any political ideology or economic interests, and there is no relationship between the different beliefs.  A belief dynamic grounded in mistrust based on different beliefs is sufficient to account for epistemic factionalization.

That said, it is important to point out that we have made several choices in our model that might have been otherwise.  We assumed that in looking at multiple beliefs, mistrust always increases in distance in each one.  But we might have considered different rules.  In particular, suppose that agents considered distance between beliefs in each arena, and used their closest beliefs to decide whom to trust.  Under this assumption, comparing multiple beliefs will never decrease social influence between agents, but has the potential to increase this influence.  In such a model, considering more beliefs could help truth-seekers by providing more opportunities to ground trust.

Such an assumption is actually closer to that made by \citet{axelrod1997dissemination}.  In his models, agents with no shared attributes do not influence one another at all.  Each possible attribute they might compare increases the chances that they share some variant, and thus have a positive probability of influencing each other.  On the contrary, in our model, adding a new problem tends to make it more likely that agents reach a level of distance, $d$, for which they no longer influence (or negatively influence) one another.

The point is that there are many different, plausible choices we might have made in this model.  One might even expect that in real-world scenarios, people will base their trust in others on a range of different considerations, of which shared belief is only one.  In this sense, the particular choices in our model may not always reflect reality.  Additionally, of course, these models are idealized in a number of other ways.  Agents are (semi) rational updaters.  All individuals are purely epistemic actors in that they seek the truth, rather than to influence each other for non-epistemic reasons.  Every individual communicates with every other individual with perfect fidelity.  And so on.

Given these idealizations, it bears reflecting on what these models can teach us.  There are a few important points here.  First, we see that under minimal conditions mistrust grounded in belief can, by itself, lead to factionalization.  Our agents do not engage in motivated reasoning about evidence: they do not, for example, exhibit confirmation bias, which occurs when agents update more strongly on evidence that confirms their beliefs.  They do not have special social ties.  They do not perceive their peers as insiders and outsiders.  But despite these things, they factionalize.  This tells us how little is needed to generate these sorts of factions.  One need not posit a common cause for all observed factions; trust dynamics suffice.  This suggests that there may be no underlying reason at all for why groups hold the particular clusters of belief that they do.

As we mention again, this result is especially notable because there is something reasonable about ignoring evidence generated by those you do not trust---particularly if you do not trust them on account of their past epistemic failures.  It would be irresponsible for scientists to update on evidence produced by known quacks.  And furthermore, there is something reasonable about deciding who is trustworthy by looking at their beliefs.  From my point of view, someone who has regularly come to hold beliefs that diverge from mine looks like an unreliable source of information.  In other words, the updating strategy used by our agents is defensible.  But, when used on the community level, it seriously undermines the accuracy of beliefs.

Of course, this point, too, should be taken with a grain of salt.  As noted, a limitation of the model is that we only consider epistemically pure agents, i.e., ones who seek truth, and who only share real, unbiased evidence.  In such a situation, it is always beneficial to trust evidence from other agents.  But other related models have considered scenarios where some agents attempt to spread false beliefs for pernicious reasons \citep{holman2015problem, holman2017experimentation, Weatherall+etal}. In these cases, it can actually benefit community members to try and identify these impure agents, and mistrust what is being shared by them.  The take-away, here, should not be that blind trust towards those around us is the best epistemic attitude.  The point, rather, is that mistrust based on shared beliefs can have negative group level effects when it is directed towards those who, like us, are just trying to figure out what is true about the world.

\section*{Acknowledgments}
This work is based on research supported by the National Science Foundation under grant \#1535139 (O'Connor).  Many thanks to anonymous referees for feedback. And thanks to the audience at the Agent-Based Models in Philosophy conference for feedback.

\end{document}